\documentclass[runningheads]{llncs}
\usepackage{graphicx}
\usepackage{hyperref}
\usepackage{color, multirow, amsmath, amssymb, tabu, tabularx, booktabs, array}
\usepackage{subfigure}
\usepackage[table,xcdraw]{xcolor}

\begin{document}
\title{BitNet: Learning-Based Bit-Depth Expansion}
\titlerunning{BitNet}

\newcommand\blfootnote[1]{%
    \begingroup
    \renewcommand\thefootnote{}\footnote{#1}%
    \addtocounter{footnote}{-1}%
    \endgroup
}

\author{Junyoung Byun\textsuperscript{*}\and Kyujin Shim\textsuperscript{*} \and Changick Kim}
\authorrunning{J. Byun et al.}

\institute{School of Electrical Engineering, KAIST, Daejeon, Republic of Korea\\
\email{\{bjyoung, kjshim1028, changick\}@kaist.ac.kr}}
\maketitle

\begin{abstract}
Bit-depth is the number of bits for each color channel of a pixel in an image. Although many modern displays support unprecedented higher bit-depth to show more realistic and natural colors with a high dynamic range, most media sources are still in bit-depth of 8 or lower. Since insufficient bit-depth may generate annoying false contours or lose detailed visual appearance, bit-depth expansion (BDE) from low bit-depth (LBD) images to high bit-depth (HBD) images becomes more and more important. In this paper, we adopt a learning-based approach for BDE and propose a novel CNN-based bit-depth expansion network (BitNet) that can effectively remove false contours and restore visual details at the same time. We have carefully designed our BitNet based on an encoder-decoder architecture with dilated convolutions and a novel multi-scale feature integration. We have performed various experiments with four different datasets including MIT-Adobe FiveK, Kodak, ESPL v2, and TESTIMAGES, and our proposed BitNet has achieved state-of-the-art performance in terms of PSNR and SSIM among other existing BDE methods and famous CNN-based image processing networks. Unlike previous methods that separately process each color channel, we treat all RGB channels at once and have greatly improved color restoration. In addition, our network has shown the fastest computational speed in near real-time.
\end{abstract}

\section{Introduction}
\blfootnote{*Authors contributed equally, listed alphabetically.}The human visual system can perceive a wide range of luminance up to 12$\sim$14 orders in magnitude \cite{percept}. Therefore modern displays have been evolved to support high dynamic range (HDR), which is a wide scope of expressable luminance, to show more natural appearance to the human eyes. Accordingly, they also have been improved to support the high bit-depth (HBD) which is essential for the HDR \cite{HDRdisplay} so that they can express more realistic and accurate colors. The bit-depth is the number of bits for each channel of a pixel and determines the total number of describable colors. Although many TVs and mobile devices such as Galaxy S9 and iPhone X already support bit-depth of 10 to satisfy HDR standards \cite{mobileHDR}, most media sources are still at bit-depth of 8 or lower, thus the displays cannot take full advantage regardless of their ability. Hence, bit-depth expansion (BDE), which converts low bit-depth (LBD) images to HBD images, has recently drawn much attention.

\begin{figure}[t]
    \label{fig:one}
    \centering
    \subfigure[8-bit image]{\includegraphics[width=40mm]{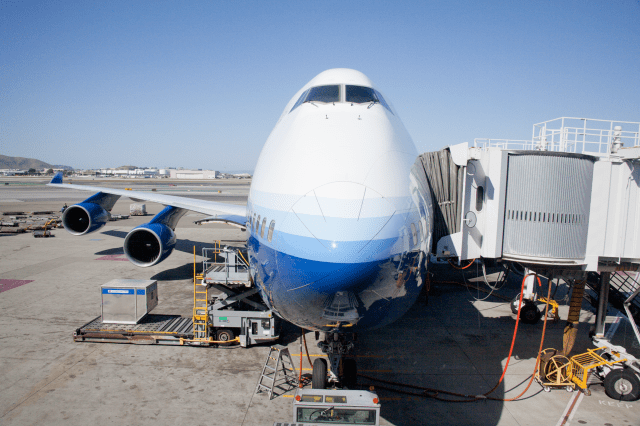}}
    \subfigure[3-bit image]{\includegraphics[width=40mm]{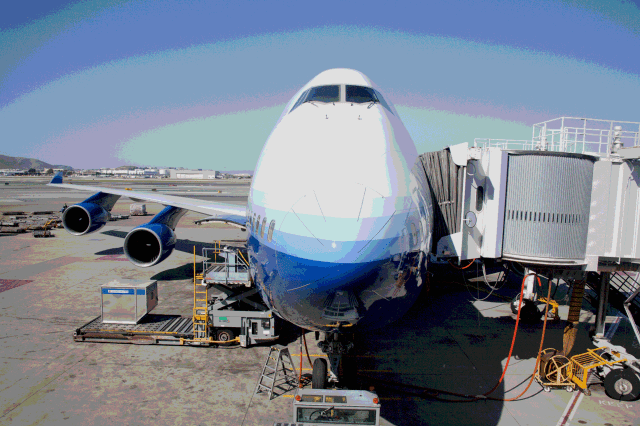}}
    \subfigure[3-bit to 8-bit result]{\includegraphics[width=40mm]{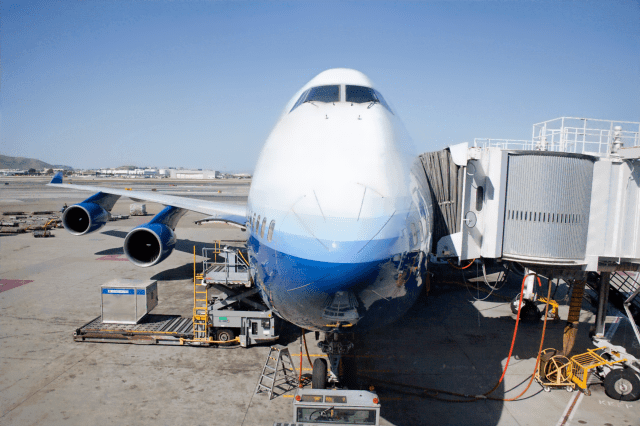}}
    \caption{An example of bit-depth expansion (BDE). In (b), false contour artifacts have occurred, and details are lost due to the insufficient bit-depth. In (c), our proposed BitNet removes false contour artifacts and restores disappeared details effectively}
\end{figure}

BDE is also useful in many other applications. It can be used as post-processing in remote control systems because they optionally reduce the bit-depth of screen data to decrease the amount of transmission for real-time operation. Additionally, it can be used as pre-processing for the professional image adjustment to prevent unwanted artifacts \cite{constrained}.

As shown in Fig. \ref{fig:one}, there are two main challenges in BDE: removing false contour artifacts and restoring disappeared visual information such as color and texture. Most existing methods \cite{adaptive,falsecontournet,noisycontour,decontouring,ACDC,CRR,CA,IPAD} only focus on eliminating false contours and are nearly unable to restore original colors. Furthermore, it is almost impossible to use them in practice because of their exceedingly long iterative operations.

To tackle these problems, we propose a carefully designed bit-depth expansion network (BitNet) based on a convolutional neural network (CNN). To the best of our knowledge, CNN-based approaches for BDE have hardly been developed. Specifically, our BitNet can precisely detect and selectively remove false contour artifacts while maintaining original edges. Furthermore, it can capture the contextual information and recover visual appearance effectively. To show this, we have demonstrated various experiments, and our network has outperformed recent BDE methods \cite{ACDC,CA,IPAD} and other famous image processing networks \cite{LRNN,fip} which are newly trained for BDE.

Our contributions are summarized as follows:

(1) We have carefully designed a network for BDE with various experiments on the effect of dilated convolutions and a novel multi-scale feature integration. Our algorithm learns inherent characteristics of natural images and eliminates false contours and restores visual details at the same time, whereas other existing methods \cite{adaptive,falsecontournet,noisycontour,decontouring,ACDC,CRR,CA,IPAD} only focus on false contour removal. As a result, we have achieved the best performance in terms of PSNR and SSIM \cite{SSIM} against other existing algorithms and newly trained famous CNN-based image processing networks. 

(2) We have dramatically increased the processing speed to enable practical application of BDE. While traditional methods take more than 20 seconds per image, our BitNet can process 768$\times$512 images at a real-time speed over 25fps.

(3) Unlike previous methods separately process each color channel, we treat all RGB channels at once. For a more detailed investigation, we have compared our network with a channel-wise version of BitNet (BitNet-Chan) and found that treating all RGB channels together makes an important improvement for accurate color restoration.

(4) We have verified that our BitNet is universally well-suited for various images with different sizes, bit-depth, and domain by using four different test datasets while it is trained with only one dataset.

This paper is organized as follows. In Section 2, we review various previous methods related to BDE. Section 3 describes the structure of our proposed network in detail. In Section 4, we describe our experiments and results. In Section 5, we examine the performance of each different structural design of our network. Section 6 concludes our paper and suggests possible future work.

\section{Related Work}
Since bit-depth expansion (BDE) became increasingly important, various methods have been proposed. The zero padding (ZP) algorithm, which is the simplest method, appends `0's after the least significant bit (LSB) of a low bit-depth (LBD) pixel. The multiplication by ideal gain (MIG) algorithm matches the scope of the current LBD space to the scope of the target high bit-depth (HBD) space by naive multiplication. The bit replication (BR) \cite{bitrep} repeatedly copies and appends current LBD bit values after LSB. Although they are fast and straightforward, they do not consider any spatial characteristics of the image. Therefore, they lose local consistency and generate strong false contour artifacts.

To deal with those problems, filter-based BDE methods have been suggested. The adaptive bit-depth expansion (ABDE) method \cite{adaptive} expands the bit-depth with ZP and segments false contour regions. Then it reduces false contours with low pass filtering (LPF). On the other hand, Park \emph{et~al.} \cite{falsecontournet} have proposed a method that detects false contours with directional contrast features and smooths them with a 3-layer fully connected network and bi-directional smoothing. However, they usually suffer from inaccurate smoothing which may also blur the true edges even with the segmentation or detection process.

At the same time, optimization-based methods have also appeared. Akira \emph{et~al.} \cite{noisycontour} have formulated the prior of target HBD values and the likelihood of noisy LBD values, and they have solved the BDE problem as a maximum a posteriori (MAP) estimation. The more advanced, ACDC algorithm \cite{ACDC} has adopted graph signal processing scheme and has formulated the BDE problem as a minimum mean squared error (MMSE) and a MAP estimation via convex optimization. It estimates HBD images from the AC signal after MAP and the DC signal after MMSE with the estimated AC signal. However, these methods are based on heuristic models and unable to account for various natural images. On the contrary, we allow the network to learn various inherent characteristics from numerous natural images, and it shows outperforming results.

One of the other major approaches is the interpolation-based methods. The contour region reconstruction (CRR) method \cite{CRR} estimates an HBD image by linear interpolation based on an up-distance map and a down-distance map from the upward contour edges and downward contour edges, respectively. However, it cannot be applied to the regions with local extrema. Therefore, the content adaptive (CA) algorithm \cite{CA} creates inner skeletons in these problematic areas and performs additional bilateral filtering for these skeleton pixels to reduce the reconstruction error.

Different from the above methods, a novel intensity potential-based BDE algorithm has been proposed recently \cite{IPAD}. It has introduced a scalar potential field to model the complex relationship among pixels in an LBD image. It first clusters connected pixels with the same intensity in each channel and finds their effective boundaries. Then an HBD image is estimated through non-linear mapping functions from the LBD image and the potential field whose field sources are the effective boundaries.

Almost all existing BDE methods are only focusing on how to remove the false contours without blurring other original contours. Also, they usually treat false contour detection and false contour removal separately. However, our BitNet deals with both detection and removal at once as well as the visual restoration with end-to-end training. On the other hand, many deep learning-based methods have been suggested for compression artifacts removal \cite{ARCNN,One-to-many}, denoising \cite{FFDNet,UniversalDenoising}, and various image processing \cite{fip,LRNN}. However, to the best of our knowledge, CNN-based approaches for BDE have hardly been developed.

\section{Methodology}

\begin{figure}[t]
	\centering
	\includegraphics[width=\linewidth]{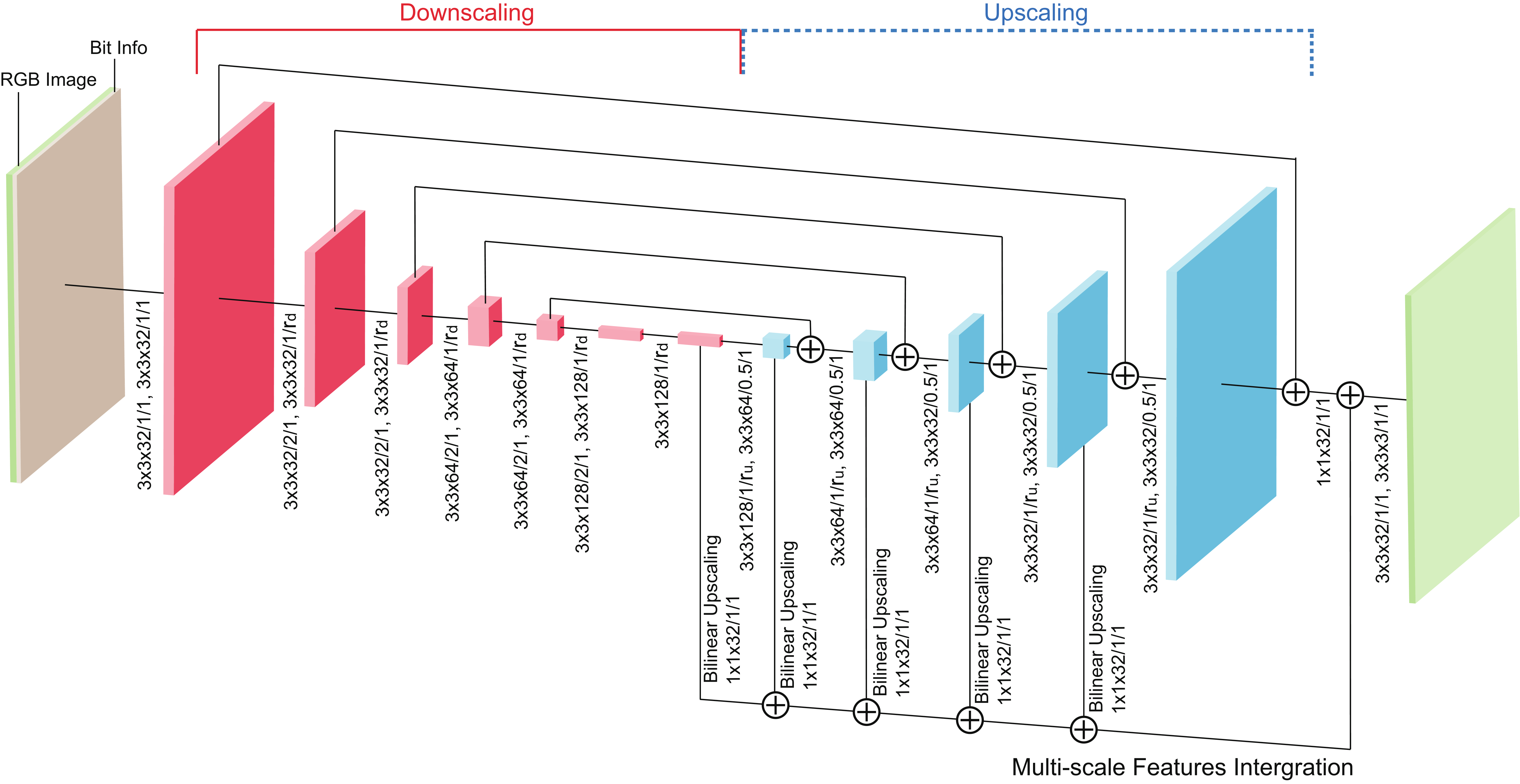}
	\caption{An illustration of our BitNet. Each block represents a feature map except for an input and a final output. The input is a concatenation of an original image and its bit-depth information. We denote the kernel size$\times$output channel/stride/dilation rate of each convolution beside the connection lines. Both $r_d$ and $r_u$ are set to two in our BitNet. Plus signs indicate the summation of corresponding feature maps.}
	\label{fig:net_archi}
\end{figure}

Our BitNet is based on an encoder-decoder structure to reduce the computational cost and carefully designed with novel architectural components. We first generate a coarse HBD image by zero padding (ZP) from an input LBD image and produce a refined HBD image through our BitNet. We illustrate the detailed network architecture in Fig. \ref{fig:net_archi}.

Our network can be divided into two parts: downscaling and upscaling. In the downscaling part, features are summarized through gradual spatial reduction. We use 3$\times$3 convolutions of stride two instead of pooling layers for a wider receptive field. In the upscaling part, we use 3$\times$3 fractionally strided convolutions \cite{frac_conv} of stride 0.5 to constantly expand the spatial size of the features until the original size. After each strided convolution for downscaling, a 3$\times$3 convolution with the dilation rate of $r_d$ is applied, while a 3$\times$3 convolution with the dilation rate of $r_u$ is performed before each strided convolution for upscaling. In our BitNet, both $r_d$ and $r_u$ are set to two for a broader receptive field which is important to handle high-resolution images.

We assume that the network would recover original colors during the downscaling and rebuild the visual appearance during the upscaling. Therefore, we add each intermediate feature from the downscaling part to each corresponding upscaled feature, as shown in Fig. \ref{fig:net_archi}, for better visual reconstruction. Also, because the receptive field is large enough, small kernels are sufficient to remove false contours and recover accurate and consistent colors even across wide and soft areas such as a cloud, lake, and wall.

We also assume that every intermediate feature would encode a resultant image with different attributes. Hence, we have designed a novel multi-scale feature integration module which merges previous features of various scales at the end of the upscaling. Specifically, we enlarge the features through bilinear interpolation to match the spatial size and employ 1$\times$1 convolutions to select appropriate features and match the number of channels. It also facilitates better gradient flow to the entire network.

To extract various features from the input image, we add two 3$\times$3 convolutional layers before downscaling. Likewise, we append 3$\times$3 and 1$\times$1 convolutional layers after the upscaling part to effectively generate the resultant HBD image from previous features. Leaky ReLU \cite{leaky_relu} pre-activation with a slope of 0.2 is applied in every convolutional layer except for the very first to maintain the dense gradient flow with enough non-linearity, and biases are added in every convolutional operation. In addition, we concatenate the bit-depth information of the original LBD image as an extra channel (i.e., an input image consists of [R, G, B, bit-depth info]) to improve the performance by identifying the bit-depth of the source and maintaining global consistency.

We have also designed a channel-wise version of BitNet (BitNet-Chan) to compare with the original version of BitNet. Every specific setting is the same as the original network except the first and last layers which are slightly modified for single-channel input and output. It processes the red, green, and blue channels separately with a single network and concatenates the results to produce the final HBD image.

\section{Experimental Results}
\subsection{Datasets}
We use MIT-Adobe FiveK dataset \cite{FiveK} to train and test our network. The dataset consists of 5,000 16-bit images taken by several photographers from a broad range of light conditions, scenes, and subjects. The dataset provides 5 sets of enhancement settings for all images made by 5 specialists. We use the photographs enhanced by expert E. Since the sizes of the images vary from 2.5 megapixels to 24.4 megapixels, the longer side of all images is scaled down to 1280 pixels by bilinear interpolation. We use the first 4,000 images (i.e., from a0001 to a4000) as training images and the remaining 1,000 images as test images.

Additionally, we adopt ESPL v2 dataset \cite{ESPL}, Kodak dataset \cite{kodak}, and TESTIMAGES dataset \cite{TESTIMAGES} to test the generalization ability of our network. The ESPL v2 dataset contains 25 1920$\times$1080 synthetic images with bit-depth of 8, and the Kodak dataset consists of 24 768$\times$512 natural images also with bit-depth of 8. Finally, the TESTIMAGES dataset includes 40 800$\times$800 images with bit-depth of 16. In the ESPL v2 dataset, we adopt only the pristine images without distortion. Besides, in the TESTIMAGES dataset, we use the photographs with the shifting indicator 'B01C00'.

\subsection{Training}
To generate training pairs for the network, we first quantize HBD images from a training set of the MIT-Adobe FiveK dataset into LBD images. Then the network is trained to minimize $l_1$ loss which is defined as
\begin{equation}
    \label{eq1}
    \centering
    l_1(I^{q_i},I) = \frac{1}{B} \sum_i \frac{1}{N_i} || f(ZP(I^{q_i}_i)) - I_i||_1,
\end{equation}
where $f(.)$ represents our BitNet, $B$ is batch size, $ZP(.)$ is a zero padding function, $I^{q_i}_i$ represents a quantized $q_i$-bit LBD image from an $i^{th}$ original HBD image $I_i$, and $N_i$ is the number of pixels in $I_i$. This $l_1$ loss minimizes difference between the resultant HBD images from BitNet and the original HBD images.

All the 3$\times$3 convolutional kernels in our BitNet are initialized by filling the center with 1 and the rest with 0 to help stable training. For the 1$\times$1 convolution layers, Xavier initialization\cite{Xavier} is applied to give sufficient variability on features aggregation. We train the network with a batch size of 1 for 100 epochs with Adam-optimizer \cite{adam}. Learning rate is set to 1e-4 for the first 75 epochs and 1e-5 for the remaining 25 epochs. Besides, we adopt three kinds of data augmentation: random horizontal flipping, random scaling from 0.5$\times$ to 1$\times$, and random bit-depth from 3 to 6. The random seed is fixed at 10,000 for every training. 

To compare with other deep learning-based methods, we also train two well-known image processing networks including CAN32+AN \cite{fip} and LRNN \cite{LRNN}. We used the official implementation\footnote{https://github.com/CQFIO/FastImageProcessing} for CAN32+AN and ported LRNN into the Tensorflow framework based on the Keras \cite{Keras} implementation\footnote{https://github.com/silverneko/Linear-RNN}. Both methods equally follow our training scheme including the additional bit-depth information channel, but LRNN has been trained with five 288$\times$288 randomly cropped images similar to the \cite{LRNN}, while three data augmentations are equally applied. Unless we do so, it takes too long to train with full-sized images. We use the cascade integration scheme for LRNN which is more appropriate for smoothing false contours. We implemented networks using Tensorflow framework \cite{tensorflow}. We also compare with other existing methods including CA \cite{CA}, ACDC \cite{ACDC}, and IPAD \cite{IPAD}, which are implemented\footnote{https://sites.google.com/site/jingliu198810/publication} in MATLAB by the author of \cite{IPAD}.

\subsection{Experiments}
We compared our BitNet with IPAD, CAN32+AN, and LRNN on the MIT-Adobe FiveK dataset. We also tested on ESPL v2, Kodak, and TESTIMAGES dataset without any additional training to compare the generalization ability. In this case, five methods including ACDC, CA, IPAD, LRNN, and CAN32+AN are compared to our BitNet in terms of average BDE quality and execution time. We performed experiments in the following system environment: Intel i7-8700k CPU, 16GB RAM, and single NVIDIA GeForce GTX 1080 Ti GPU.

In addition to the figures in the paper, more results can be seen in the supplementary material. Note that all images in the paper and the supplementary material are set as bit-depth of 8 for display. Also, we recommend viewing the images in a proper electronic form for their clear appearance and recognition.

\subsection{Qualitative Results}
\begin{figure}[t!]
    \centering
    \newcommand\w{0.2\linewidth}
    \begin{tabular}{cccc}        \raisebox{0.25\height}{\includegraphics[width=\w,height=\w,keepaspectratio]{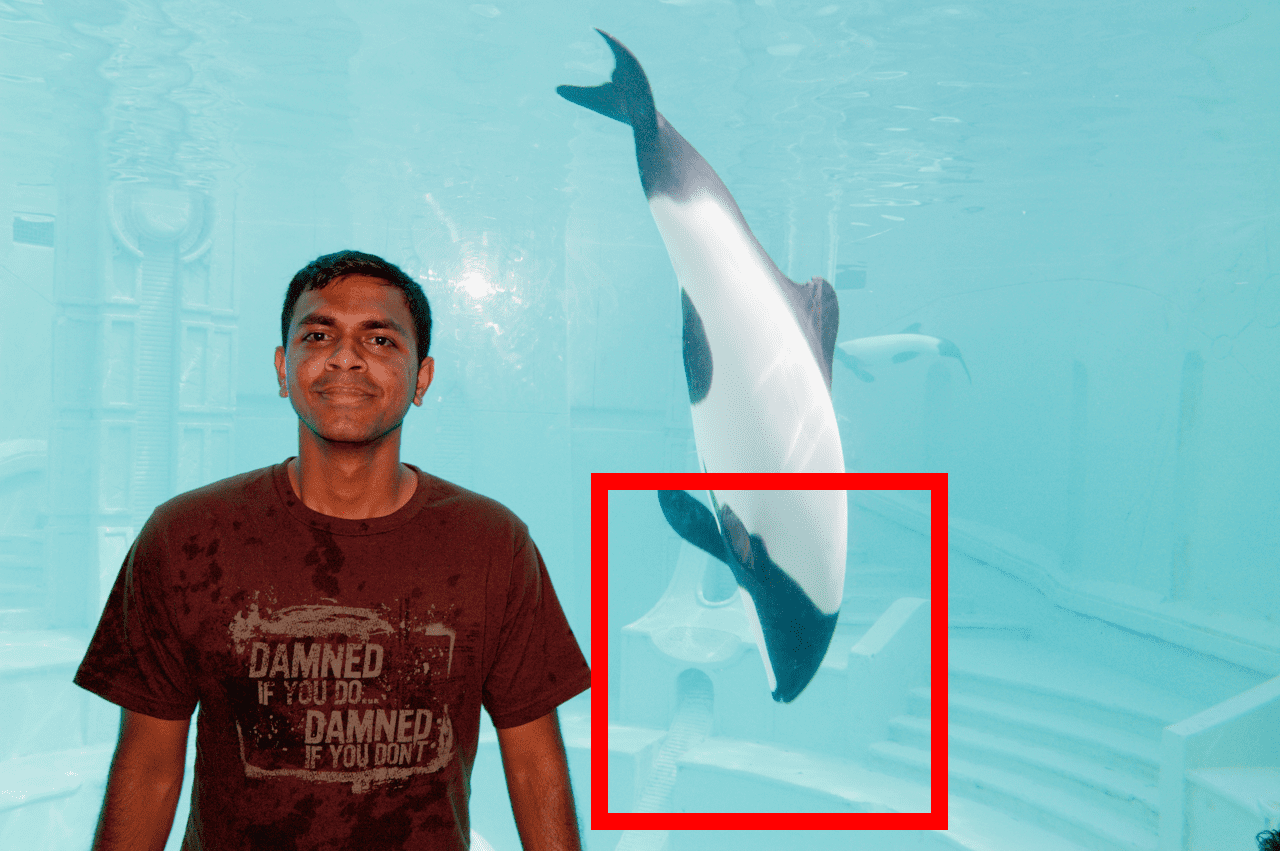}} &
        \includegraphics[width=\w]{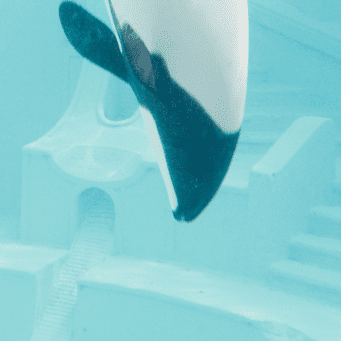} &
        \includegraphics[width=\w]{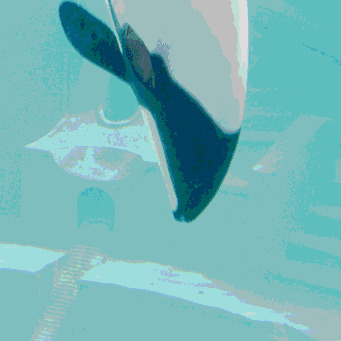} &
        \includegraphics[width=\w]{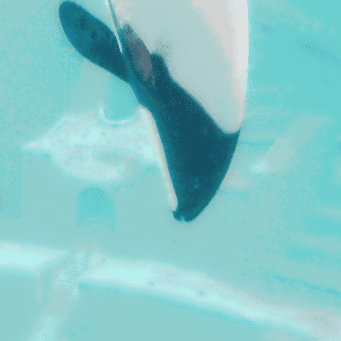} \\
        \includegraphics[width=\w,height=\w,keepaspectratio]{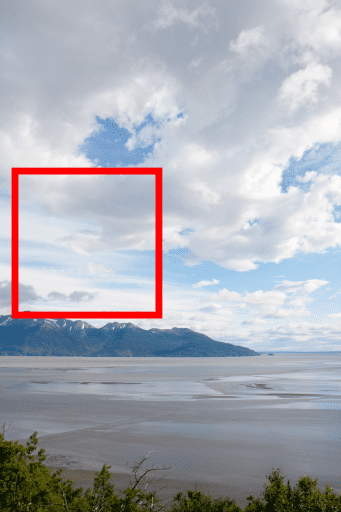} &
        \includegraphics[width=\w]{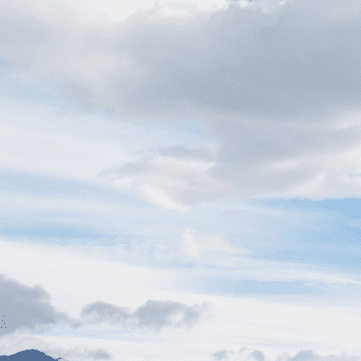} &
        \includegraphics[width=\w]{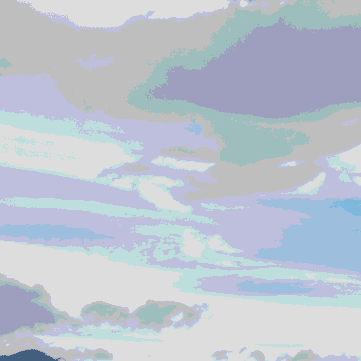} &
        \includegraphics[width=\w]{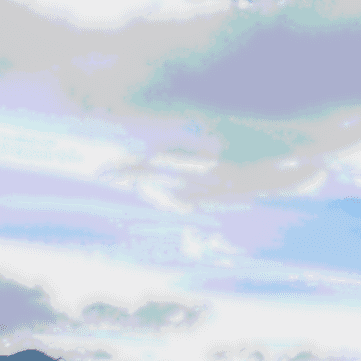} \\
        \includegraphics[width=\w,height=\w,keepaspectratio]{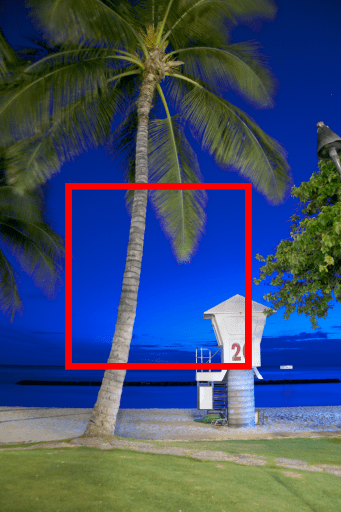}  &
        \includegraphics[width=\w]{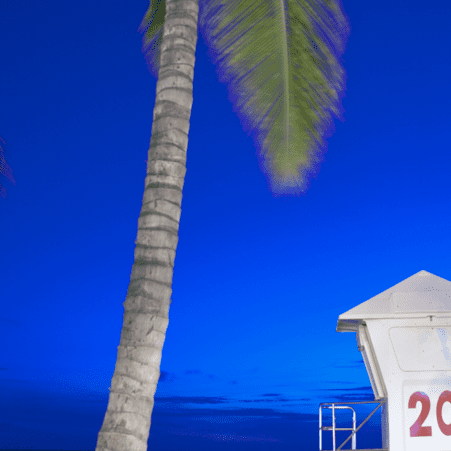} &
        \includegraphics[width=\w]{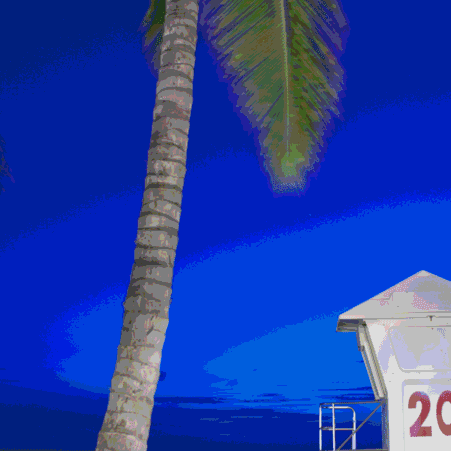} &
        \includegraphics[width=\w]{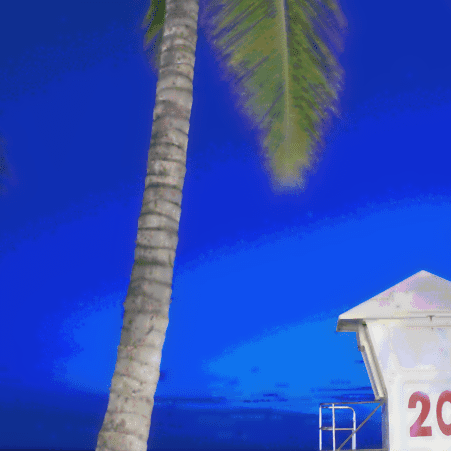} \\
        (a) Original & (b) Cropped  & (c) Input  & (d) IPAD\\ \addlinespace[1mm]
        \includegraphics[width=\w]{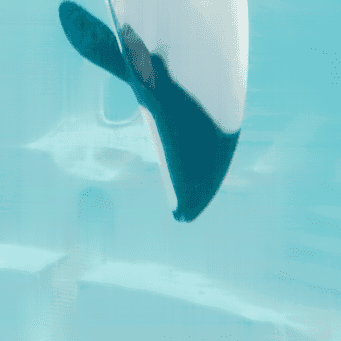} &
        \includegraphics[width=\w]{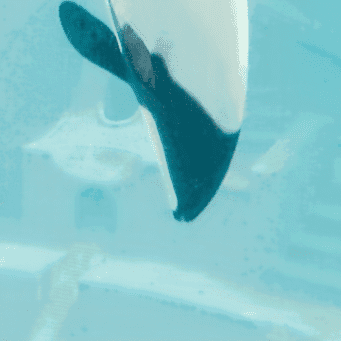} &
        \includegraphics[width=\w]{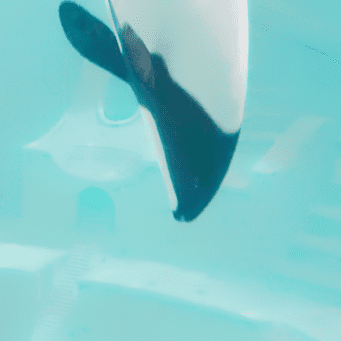} &
        \includegraphics[width=\w]{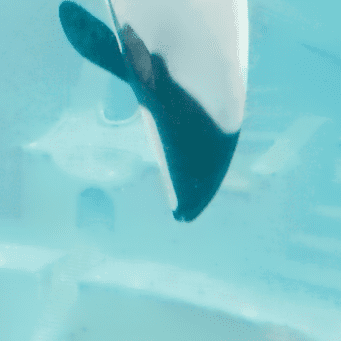} \\
        \includegraphics[width=\w]{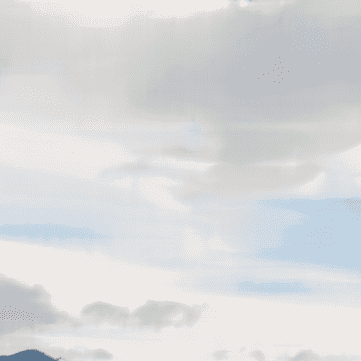} &
        \includegraphics[width=\w]{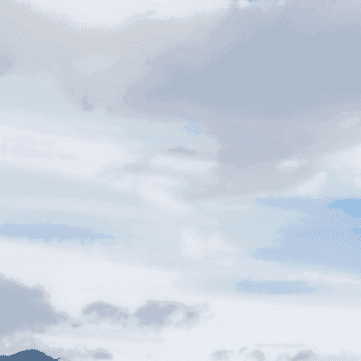} &
        \includegraphics[width=\w]{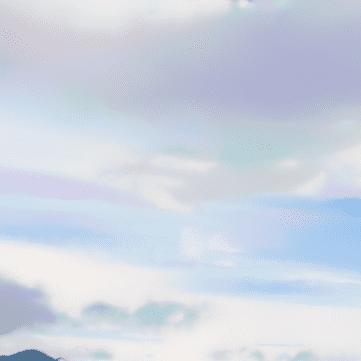} &
        \includegraphics[width=\w]{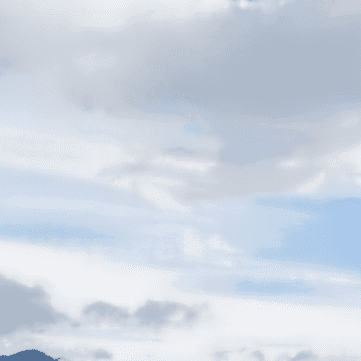} \\
        \includegraphics[width=\w]{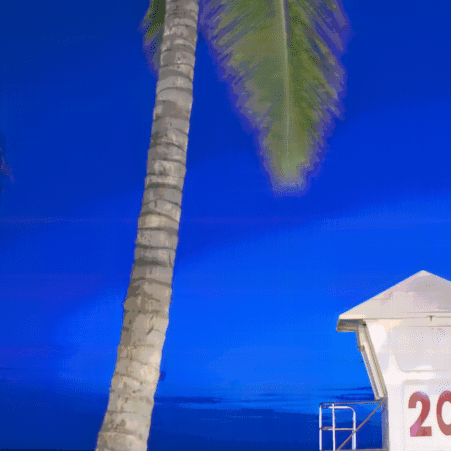} &
        \includegraphics[width=\w]{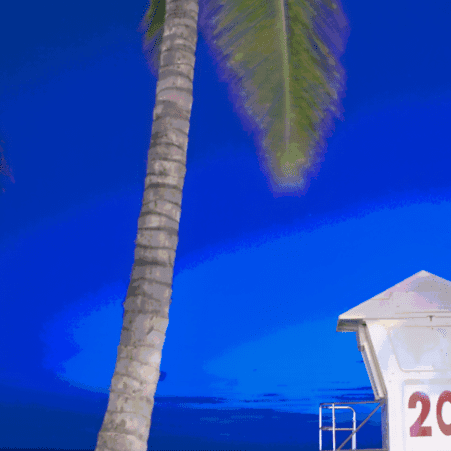} &
        \includegraphics[width=\w]{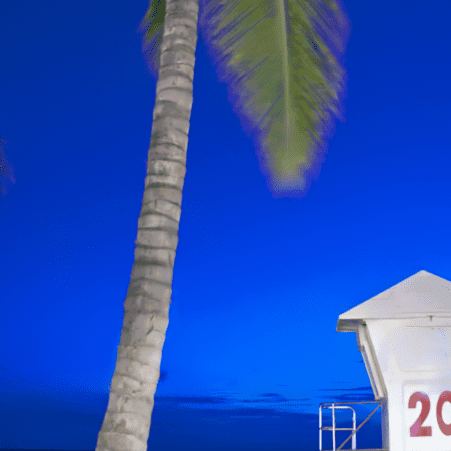} &
        \includegraphics[width=\w]{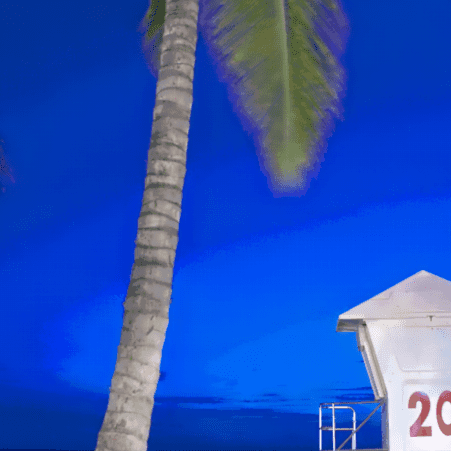} \\
        (e) LRNN & (f) CAN32+AN & (g) BitNet-Chan  & (h) BitNet
        \end{tabular}
    \caption{Qualitative comparisons of 3-bit to 8-bit results with MIT-Adobe FiveK}
    \label{fig:multi1}
\end{figure}

Figure \ref{fig:multi1} shows the test results of expansion from 3-bits to 8-bits using IPAD and four different deep networks on the MIT-Adobe FiveK dataset \cite{FiveK}. The first image of Fig. \ref{fig:multi1}(h) clearly shows that BitNet well restores underwater structural details. In the second image which includes the clouds, BitNet effectively removes false contours and also reverts the erroneous colors to natural colors. On the contrary, IPAD and BitNet-Chan have difficulties in restoring natural color tone. They handle each RGB channel separately, so they do not have enough information to effectively restore the colors. However, the third image with the night sky shows that Bit-Chan removes false contours much better. This means the smoothing ability of BitNet sharply drops on unseen color patterns, while Bit-Chan well removes false contours regardless of the colors.

\begin{figure}[t!]
    \centering
    \newcommand\sw{0.19\linewidth}
    \begin{tabular}{ccccc}
        \raisebox{0.3\height}{\includegraphics[width=\sw,height=\sw,keepaspectratio]{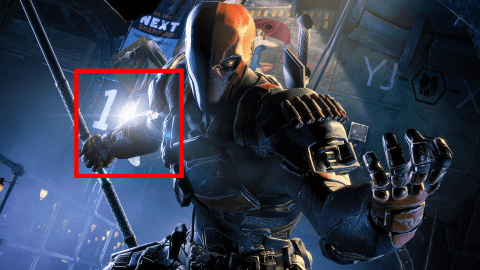}} &
        \includegraphics[width=\sw]{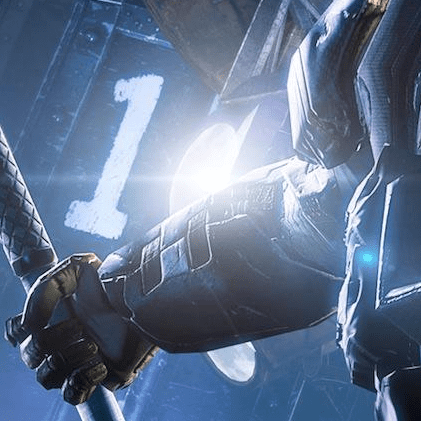} &
        \includegraphics[width=\sw]{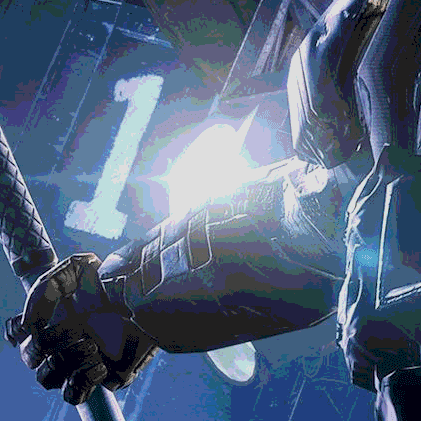} &
        \includegraphics[width=\sw]{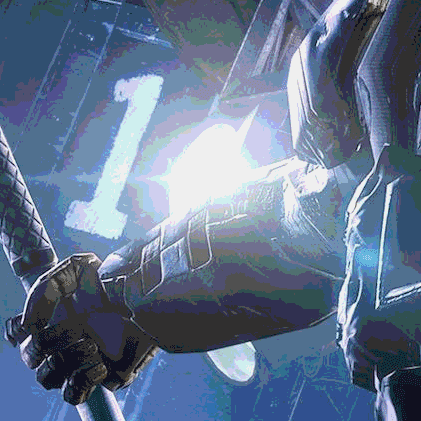} &
        \includegraphics[width=\sw]{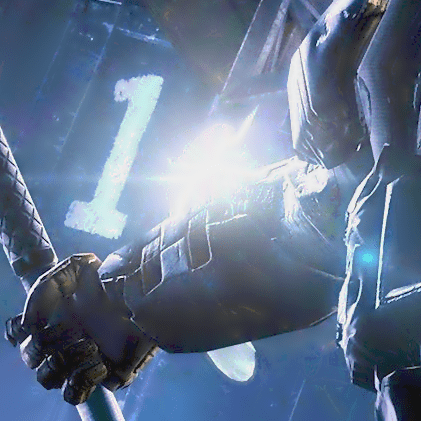} \\ \includegraphics[width=\sw,height=\sw,keepaspectratio]{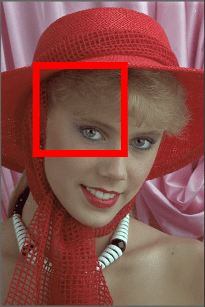} &
        \includegraphics[width=\sw]{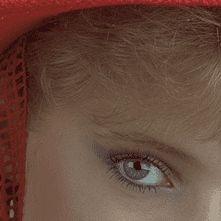} &
        \includegraphics[width=\sw]{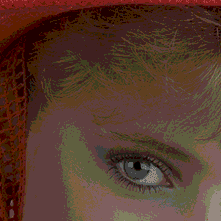} &
        \includegraphics[width=\sw]{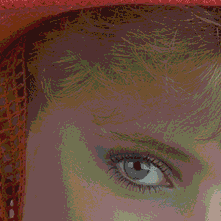} &
        \includegraphics[width=\sw]{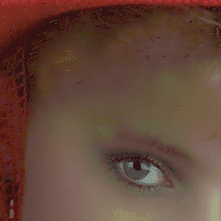} \\ \includegraphics[width=\sw,height=\sw,keepaspectratio]{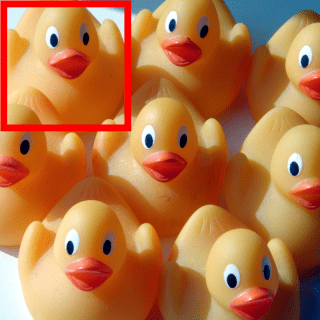} &
        \includegraphics[width=\sw]{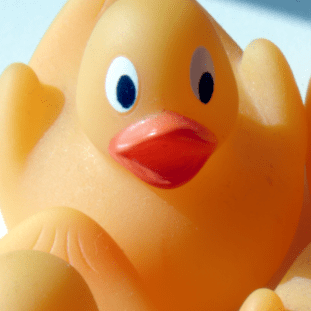} &
        \includegraphics[width=\sw]{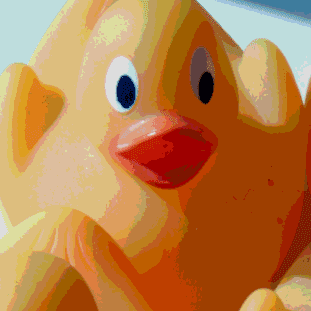} &
        \includegraphics[width=\sw]{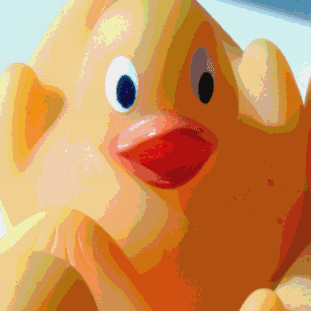} &
        \includegraphics[width=\sw]{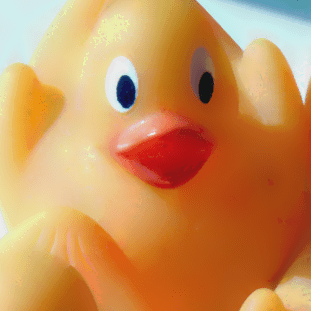} \\
        (a) Original & (b) Cropped & (c) Input & (d) ACDC & (e) CA\\
        \addlinespace[1mm]
        \includegraphics[width=\sw]{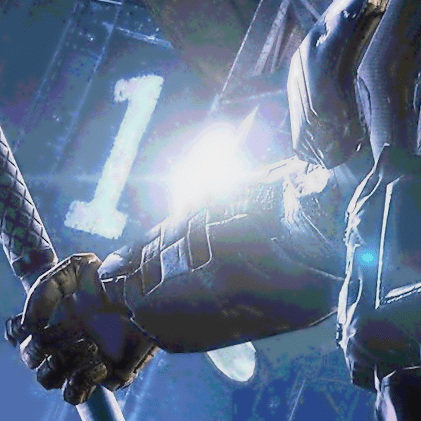} &
        \includegraphics[width=\sw]{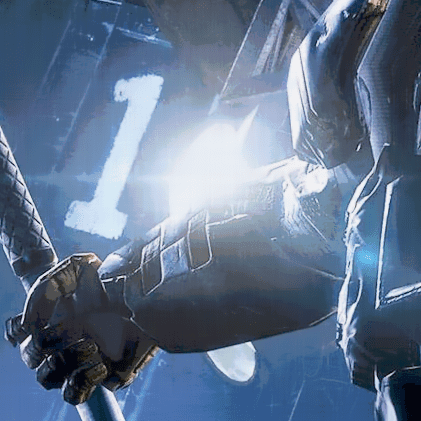} &
        \includegraphics[width=\sw]{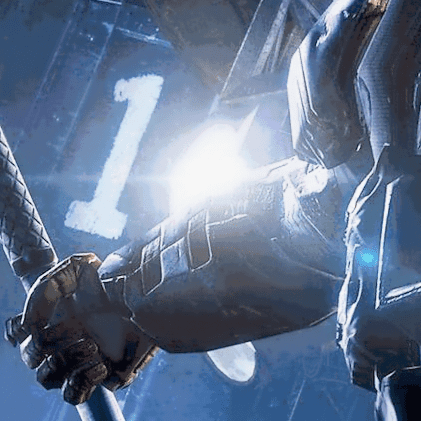} &
        \includegraphics[width=\sw]{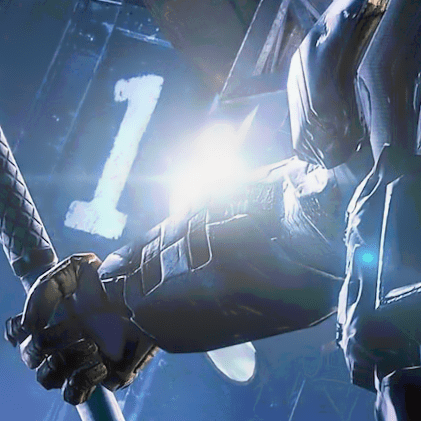} &
        \includegraphics[width=\sw]{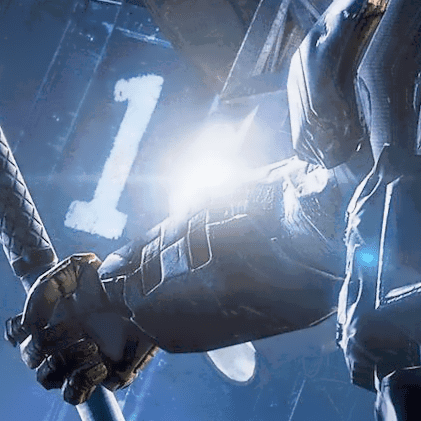}\\
        \includegraphics[width=\sw]{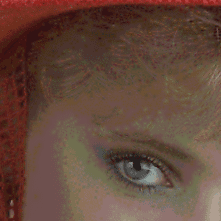} &
        \includegraphics[width=\sw]{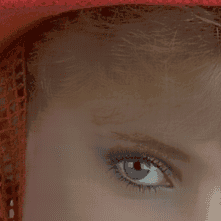} &
        \includegraphics[width=\sw]{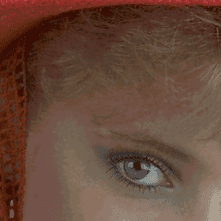} &
        \includegraphics[width=\sw]{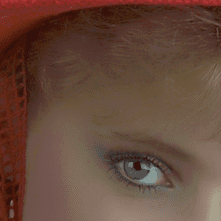} &
        \includegraphics[width=\sw]{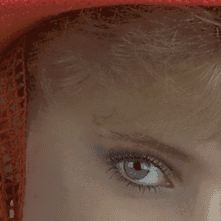}\\
        \includegraphics[width=\sw]{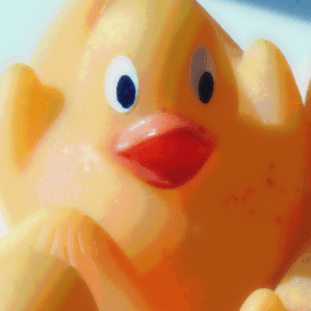} &
        \includegraphics[width=\sw]{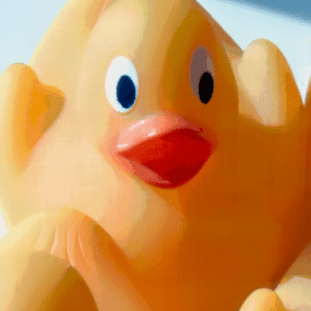} &
        \includegraphics[width=\sw]{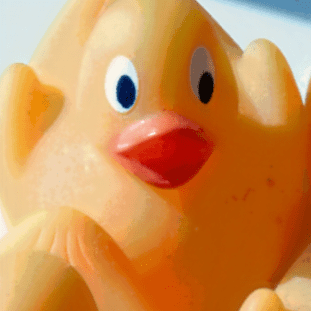} &
        \includegraphics[width=\sw]{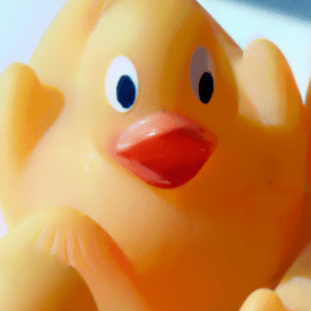} &
        \includegraphics[width=\sw]{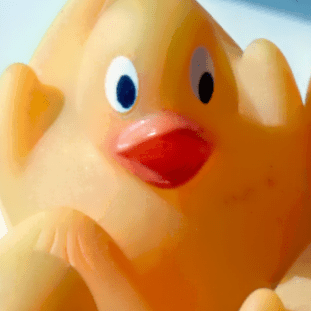}\\
        (f) IPAD & (g) LRNN & (h) CAN32+AN & (i) BitNet-Chan & (j) BitNet
    \end{tabular}
    \caption{Qualitative comparisons of 3-bit to 8-bit results. From the top, three images are from ESPL v2, Kodak, and TESTIMAGES dataset, respectively}
    \label{fig:multi2}
\end{figure}

Deep convolutional networks may overfit on the dataset used for training. Therefore, we tested on the three different datasets including well-known ESPL v2 dataset \cite{ESPL}, Kodak dataset \cite{kodak}, and TESTIMAGES dataset \cite{TESTIMAGES} without any additional training to verify the generalization ability of each. Figure \ref{fig:multi2} shows the results with three images from ESPL v2, Kodak, and TESTIMAGES dataset, respectively. BitNet demonstrates its outstanding ability for false contour removal and color restoration.
In the first image, BitNet better restores the original color of the heavy armor and express the light source more naturally. In addition, BitNet well eliminates the false contours in the second photograph while maintaining the details of the eyelashes and the hair. On the other hand, BitNet-Chan has demonstrated remarkable smoothing capabilities in the duck image while CA \cite{CA} and ACDC \cite{ACDC} yield glow artifacts.

\subsection{Quantitative Results}
\begin{table}[t!]
    \centering
    \setlength{\tabcolsep}{2pt}
    \setlength{\arrayrulewidth}{.13em}
    \renewcommand{\arraystretch}{1.25}
    \newcolumntype{?}{!{\vrule width .13em}}
    \caption{Average evaluation results. The best result is highlighted in bold.}
    \label{table:comparison}
    \begin{tabular}{c?cccccccc}
        \hline
        & \multicolumn{8}{c}{\bf{MIT-Adobe FiveK dataset \cite{FiveK}}} \\
        \cline{2-9}
        & \multicolumn{2}{c}{3 $\rightarrow$ 16} & \multicolumn{2}{c}{4 $\rightarrow$ 16} & \multicolumn{2}{c}{5 $\rightarrow$ 16} & \multicolumn{2}{c}{6 $\rightarrow$ 16} \\
        \cline{2-9}
        & PSNR & SSIM & PSNR & SSIM & PSNR & SSIM & PSNR & SSIM \\
        \hline
        IPAD     \cite{IPAD} & 29.283      & 0.8425      & 34.990      & 0.9201      & 40.404      & 0.9628      & 45.787      & 0.9847      \\
        LRNN     \cite{LRNN} & 31.422      & 0.8884      & 35.876      & 0.9454      & 37.679      & 0.9691      & 38.598      & 0.9787      \\
        CAN32+AN \cite{fip}  & 32.277      & 0.8964      & 37.781      & 0.9537      & 42.642      & 0.9809      & 47.042      & 0.9926      \\
        BitNet-Chan          & 33.162      & 0.8993      & 38.581      & 0.9565      & \bf{43.424} & 0.9825      & \bf{48.225} & \bf{0.9935} \\
        BitNet               & \bf{33.426} & \bf{0.9086} & \bf{38.634} & \bf{0.9587} & 43.291      & \bf{0.9826} & 47.826      & 0.9933      \\
        \hline
        & \multicolumn{4}{c}{\bf{ESPL v2 dataset \cite{ESPL}}} & \multicolumn{4}{c}{\bf{Kodak dataset \cite{kodak}}} \\
        \cline{2-9}
        & \multicolumn{2}{c}{3 $\rightarrow$ 8} & \multicolumn{2}{c}{4 $\rightarrow$ 8} & \multicolumn{2}{c}{3 $\rightarrow$ 8} & \multicolumn{2}{c}{4 $\rightarrow$ 8} \\
        \cline{2-9}
        & PSNR & SSIM & PSNR & SSIM & PSNR & SSIM & PSNR & SSIM \\
        \hline
        ACDC     \cite{ACDC} & 28.594      & 0.7766      & 34.329      & 0.8756      & 28.656      & 0.8196      & 34.682      & 0.9132      \\
        CA       \cite{CA}   & 29.375      & 0.8270      & 35.511      & 0.9171      & 29.145      & 0.8439      & 34.738      & 0.9312      \\
        IPAD     \cite{IPAD} & 29.776      & 0.8381      & 35.762      & 0.9186      & 29.201      & 0.8536      & 34.908      & 0.9340      \\
        LRNN     \cite{LRNN} & 31.377      & 0.8480      & 37.094      & 0.9200      & 31.368      & 0.8984      & 36.615      & 0.9559      \\
        CAN32+AN \cite{fip}  & 31.645      & 0.8633      & 37.870      & 0.9320      & 31.778      & 0.9054      & 38.037      & 0.9620      \\
        BitNet-Chan          & 31.975      & 0.8469      & 38.401      & 0.9381      & 31.594      & 0.8886      & 37.479      & 0.9547      \\
        BitNet               & \bf{32.876} & \bf{0.8802} & \bf{38.968} & \bf{0.9459} & \bf{32.683} & \bf{0.9181} & \bf{38.482} & \bf{0.9657} \\
        \hline
        & \multicolumn{8}{c}{\bf{TESTIMAGES dataset \cite{TESTIMAGES}}} \\
        \cline{2-9}
        & \multicolumn{2}{c}{3 $\rightarrow$ 16} & \multicolumn{2}{c}{4 $\rightarrow$ 16} & \multicolumn{2}{c}{5 $\rightarrow$ 16} & \multicolumn{2}{c}{6 $\rightarrow$ 16} \\
        \cline{2-9}
        & PSNR & SSIM & PSNR & SSIM & PSNR & SSIM & PSNR & SSIM \\
        \hline
        ACDC     \cite{ACDC} & 28.800      & 0.8207      & 34.763      & 0.9069      & 40.769      & 0.9634      & 46.788      & 0.9883      \\
        CA       \cite{CA}   & 28.886      & 0.8655      & 34.821      & 0.9343      & 40.173      & 0.9722      & 45.010      & 0.9906      \\
        IPAD     \cite{IPAD} & 29.494      & 0.8770      & 35.573      & 0.9414      & 41.188      & 0.9738      & 46.610      & 0.9895      \\
        LRNN     \cite{LRNN} & 31.321      & 0.8912      & 37.108      & 0.9484      & 40.153      & 0.9743      & 42.604      & 0.9865      \\
        CAN32+AN \cite{fip}  & 31.215      & 0.8929      & 37.584      & 0.9525      & 42.945      & 0.9810      & 47.543      & 0.9924      \\
        BitNet-Chan          & \bf{32.598} & 0.9064      & \bf{38.956} & \bf{0.9630} & \bf{44.481} & \bf{0.9857} & \bf{49.404} & \bf{0.9948} \\
        BitNet               & 32.136      & \bf{0.9069} & 38.337      & 0.9584      & 43.791      & 0.9836      & 48.672      & 0.9941      \\
        \hline
    \end{tabular}
\end{table}
For quantitative evaluation, we measured the peak signal to noise ratio (PSNR) and the structural similarity (SSIM) \cite{SSIM} between the original HBD image and the resultant HBD image, following the previous work \cite{CA,IPAD}.

\begin{table}[t!]
    \centering
    \setlength{\tabcolsep}{2pt}

    \setlength{\arrayrulewidth}{.13em}
    \newcolumntype{?}{!{\vrule width .13em}}
    \caption{Average execution time per image. ESPL v2 contains 1920$\times$1080 images, Kodak consists of 768$\times$512 images, and TESTIMAGES includes 800$\times$800 images. Unit is second for all. The best result is highlighted in bold.}
    \label{table:speed}
    \renewcommand{\arraystretch}{1.25}
    \begin{tabular}{c?cccccccc}
        \hline
        & \multicolumn{4}{c}{\bf{ESPL v2 dataset \cite{ESPL}}} & \multicolumn{4}{c}{\bf{Kodak dataset \cite{kodak}}} \\
        \cline{2-9}
        & \multicolumn{2}{c}{3 $\rightarrow$ 8} & \multicolumn{2}{c}{4 $\rightarrow$ 8} & \multicolumn{2}{c}{3 $\rightarrow$ 8} & \multicolumn{2}{c}{4 $\rightarrow$ 8} \\
        \cline{2-9}
        & GPU & CPU & GPU & CPU & GPU & CPU & GPU & CPU \\
        \hline
        ACDC     \cite{ACDC} & -          & 5203.395   & -          & 5382.344   & -          & 934.351    & -          & 924.973    \\
        CA       \cite{CA}   & -          & 737.851    & -          & 520.057    & -          & 147.869    & -          & 142.835    \\
        IPAD     \cite{IPAD} & -          & 339.107    & -          & 280.251    & -          & 33.833     & -          & 22.908     \\
        LRNN     \cite{LRNN} & 0.874      & 4.882      & 0.875      & 4.871      & 0.312      & 0.952      & 0.312      & 0.943      \\
        CAN32+AN \cite{fip}  & 0.367      & 7.140      & 0.366      & 7.161      & 0.071      & 1.361      & 0.071      & 1.348      \\
        BitNet-Chan          & 0.585      & 8.944      & 0.586      & 8.938      & 0.113      & 1.719      & 0.113      & 1.667      \\
        BitNet               & \bf{0.198} & \bf{2.976} & \bf{0.198} & \bf{2.957} & \bf{0.038} & \bf{0.561} & \bf{0.038} & \bf{0.554} \\
        \hline
        & \multicolumn{8}{c}{\bf{TESTIMAGES dataset \cite{TESTIMAGES}}} \\
        \cline{2-9}
        & \multicolumn{2}{c}{3 $\rightarrow$ 16} & \multicolumn{2}{c}{4 $\rightarrow$ 16} & \multicolumn{2}{c}{5 $\rightarrow$ 16} & \multicolumn{2}{c}{6 $\rightarrow$ 16} \\
        \cline{2-9}
        & GPU & CPU & GPU & CPU & GPU & CPU & GPU & CPU \\
        \hline 
        ACDC     \cite{ACDC} & -          & 1560.481   & -          & 1541.790  & -          & 1442.176  & -          & 1407.948  \\
        CA       \cite{CA}   & -          & 177.715    & -          & 133.238   & -          & 96.041    & -          & 63.278    \\
        IPAD     \cite{IPAD} & -          & 32.156     & -          & 33.427    & -          & 40.452    & -          & 52.538    \\
        LRNN     \cite{LRNN} & 0.386      & 1.513      & 0.390      & 1.509     & 0.390      & 1.507     & 0.395      & 1.508     \\
        CAN32+AN \cite{fip}  & 0.115      & 2.207      & 0.115      & 2.193     & 0.115      & 2.180     & 0.115      & 2.177     \\
        BitNet-Chan          & 0.182      & 2.702      & 0.161      & 2.697     & 0.182      & 2.693     & 0.182      & 2.689     \\
        BitNet               & \bf{0.062} & \bf{0.898} & \bf{0.062} &\bf{0.895} & \bf{0.062} &\bf{0.893} & \bf{0.062} &\bf{0.897} \\ 
        \hline
    \end{tabular}
\end{table}

Table \ref{table:comparison} includes the average evaluation results with the MIT-Adobe FiveK dataset. Our proposed networks outperform other existing methods in every case. Compared to IPAD, which is an existing state-of-the-art method for BDE, our method shows a significant improvement in both PSNR and SSIM. In particular, for the BDE from 3-bits to 16-bits with the MIT-Adobe FiveK dataset, the PSNR and SSIM gains are 4.143 and 0.0661, respectively. Note that BitNet shows better performance than BitNet-Chan for 3-bit and 4-bit input, whereas BitNet-Chan shows a higher PSNR for 6-bit input. However, in the case of BDE from 6-bits to 16-bits, the visual differences are very minor among the results.

The results from the other three datasets are also shown in Table \ref{table:comparison}. Compared to others, our method shows much higher performances in all cases. In the ESPL v2 dataset and the Kodak dataset, BitNet outperforms than BitNet-Chan. However, for the TESTIMAGES dataset, BitNet-Chan shows better results than BitNet. It is because BDE on the TESTIMAGES dataset mainly requires false contour removal, not color restoration, and BitNet-Chan has better smoothing capabilities than BitNet.

\subsection{Execution Time}

Table \ref{table:speed} shows the average processing time per image on the three datasets. ACDC, CA, and IPAD show different execution times depending on both image sizes and the number of bit-depth, while execution times of CNN-based methods mainly depend on the image resolution. While ACDC, CA, and IPAD take more than 20 seconds per image, BitNet shows real-time speeds of 25fps or higher for 768$\times$512 images with a single GPU. Note that our BitNet is at least 500 times faster than IPAD with a GPU, and compared to LRNN and CAN, BitNet is about 2 to 8 times faster, even with better performance. Meanwhile, BitNet-Chan is about three times slower than BitNet because it has to infer each color channel separately due to memory limitations of the GPU.

\section{Discussion}

\begin{table}[t!]
    \centering
    \setlength{\tabcolsep}{1.8pt}
    \setlength{\arrayrulewidth}{.13em}
    \renewcommand{\arraystretch}{1.25}
    \newcolumntype{?}{!{\vrule width .13em}}
    \caption{Performance comparison to investigate the effect of the bit-depth information channel and dilation rates. The numbers inside the brackets denote $\langle$$r_d$, $r_u$$\rangle$, which are the rates of convolution layers in downscaling and upscaling part, respectively. The best result is highlighted in bold.}
    \label{table:dial_bit}
    \begin{tabular}{c?cccccccc}
        \hline
        & \multicolumn{8}{c}{\bf{MIT-Adobe FiveK dataset \cite{FiveK}}} \\
        \cline{2-9}
        & \multicolumn{2}{c}{3 $\rightarrow$ 16} & \multicolumn{2}{c}{4 $\rightarrow$ 16} & \multicolumn{2}{c}{5 $\rightarrow$ 16} & \multicolumn{2}{c}{6 $\rightarrow$ 16} \\
        \cline{2-9}
        & PSNR & SSIM & PSNR & SSIM & PSNR & SSIM & PSNR & SSIM \\
        \hline
        $\langle$1, 1$\rangle$ w/o     bit info & 33.064      & 0.9043      & 38.408      & 0.9572      & 43.133      & 0.9822      & 47.665      & 0.9931      \\
        $\langle$1, 2$\rangle$ w/o     bit info & 32.934      & 0.9027      & 38.262      & 0.9564      & 42.868      & 0.9817      & 47.463      & 0.9929      \\
        $\langle$2, 1$\rangle$ w/o     bit info & 32.966      & 0.9041      & 38.321      & 0.9572      & 42.948      & 0.9821      & 47.332      & 0.9930      \\
        $\langle$2, 2$\rangle$ w/o     bit info & 33.018      & 0.9044      & 38.271      & 0.9569      & 43.012      & 0.9819      & 47.410      & 0.9929      \\
        $\langle$1, 1$\rangle$ w/\quad bit info & 33.323      & 0.9079      & 38.632      & 0.9585      & \bf{43.372} & \bf{0.9827} & \bf{47.973} & \bf{0.9934} \\
        $\langle$1, 2$\rangle$ w/\quad bit info & 33.333      & 0.9066      & 38.627      & 0.9585      & 43.369      & \bf{0.9827} & 47.946      & \bf{0.9934} \\
        $\langle$2, 1$\rangle$ w/\quad bit info & 33.361      & 0.9076      & 38.624      & 0.9581      & 43.358      & 0.9825      & 48.024      & 0.9933      \\
        $\langle$2, 2$\rangle$ w/\quad bit info & \bf{33.426} & \bf{0.9086} & \bf{38.634} & \bf{0.9587} & 43.291      & 0.9826      & 47.826      & 0.9933      \\
        \hline
    \end{tabular}
\end{table}

\subsection{Bit-Depth Information Channel}
To see the effectiveness of the bit-depth information channel, we trained the network without the additional channel to the RGB input. Table \ref{table:dial_bit} shows the results with the MIT-Adobe FiveK dataset. The bit-depth information channel identifies the bit-depth of the input to the network and improves the performance in all cases.

\subsection{Dilation Rates}

To investigate the effect of the dilation rates, we trained and tested the networks of various rates with the MIT-Adobe FiveK dataset. The dilation rates of convolution layers in downscaling and upscaling part, which are denoted as $r_d$ and $r_u$ in Fig. \ref{fig:net_archi}, were equally set in each part. We compared four different combinations (i.e $\langle$1, 1$\rangle$, $\langle$2, 2$\rangle$, $\langle$2, 1$\rangle$ and $\langle$2, 2$\rangle$) of $r_d$ and $r_u$, and Table \ref{table:dial_bit} shows the results. Without the bit-depth information channel, the dilation rates do not affect the performance much. However, when the bit-depth information is adopted, the $\langle$2, 2$\rangle$ combination shows better performance for the input with bit-depth of 3 or 4. Although its performance is slightly degraded when the input is bit-depth of 5 or 6, PSNR is already over 43, and the difference of SSIM is minuscule at 1e-4, so the visual differences are very little among the results. Therefore, we set both $r_d$ and $r_u$ as two in our BitNet, because it works better for more difficult cases and has a wider receptive field which is important to handle high-resolution images.

\subsection{Multi-scale Feature Integration}

\begin{figure}[t!]
    \centering
        \begin{tabular}{ccccc}
        \raisebox{0.2\height}{\includegraphics[width=22mm]{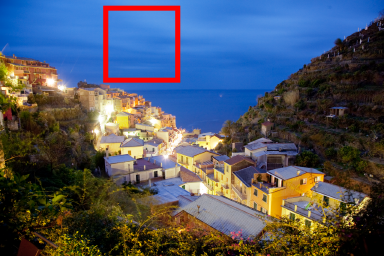}} &
        \includegraphics[width=22mm]{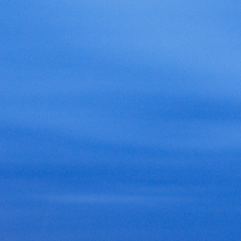} &
        \includegraphics[width=22mm]{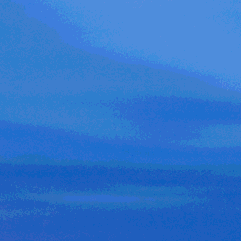} &
        \includegraphics[width=22mm]{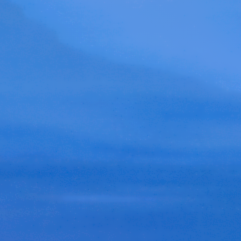} &
        \includegraphics[width=22mm]{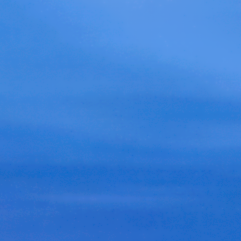} \\
        \raisebox{0.2\height}{\includegraphics[width=22mm]{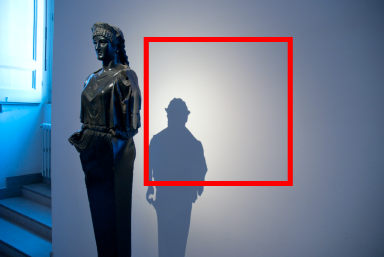}} &
        \includegraphics[width=22mm]{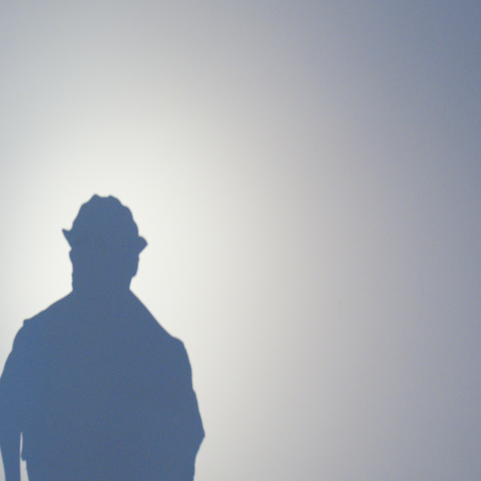} &
        \includegraphics[width=22mm]{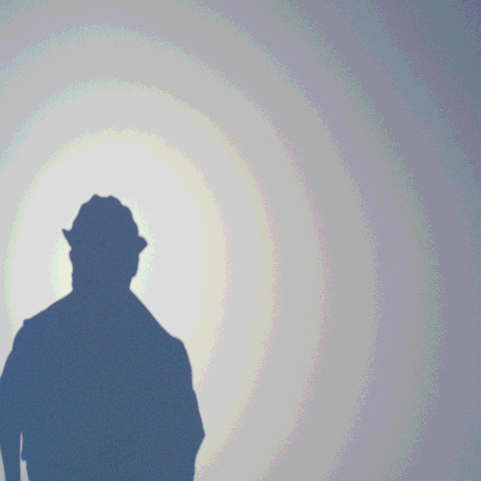} &
        \includegraphics[width=22mm]{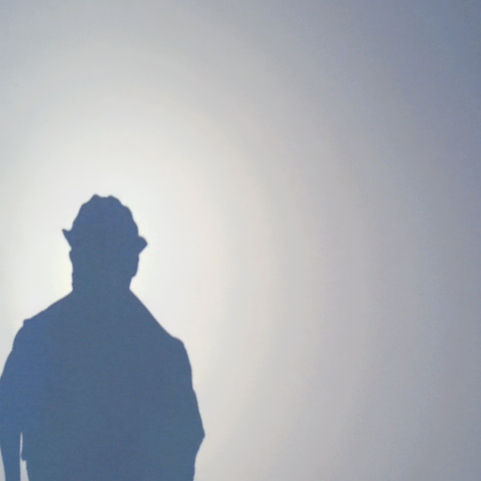} &
        \includegraphics[width=22mm]{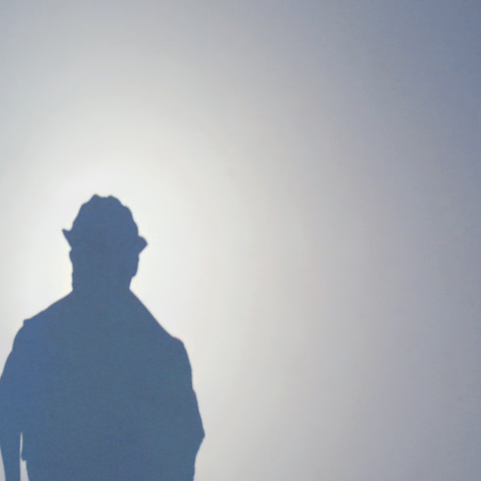} \\
        (a) 8-bit original & (b) Cropped & (c) 4-bit input & (d) w/o MSFI & (e) w/ MSFI\\
        \end{tabular}
    \caption{Comparison of 4-bit to 8-bit results w/ and w/o multi-scale feature integration}
    \label{fig:multi4}
\end{figure}

\begin{table}[t!]
    \centering
    \setlength{\tabcolsep}{1.5pt}
    \setlength{\arrayrulewidth}{.13em}
    \renewcommand{\arraystretch}{1.25}
    \newcolumntype{?}{!{\vrule width .13em}}
    \caption{Performance comparison w/ and w/o multi-scale feature integration on the MIT-Adobe FiveK Dataset. The best result is highlighted in bold.}
    \label{table:integration}
    \begin{tabular}{c?cccccccc}
        \hline
        & \multicolumn{8}{c}{\bf{MIT-Adobe FiveK dataset \cite{FiveK}}} \\
        \cline{2-9}
        & \multicolumn{2}{c}{3 $\rightarrow$ 16} & \multicolumn{2}{c}{4 $\rightarrow$ 16} 
                 & \multicolumn{2}{c}{5 $\rightarrow$ 16} & \multicolumn{2}{c}{6 $\rightarrow$ 16} \\
        \cline{2-9}
        & PSNR & SSIM & PSNR & SSIM & PSNR & SSIM & PSNR & SSIM \\
        \hline
        2, 2 w/o integration & 33.208      & 0.9072      & 38.262      & 0.9582      & \bf{43.294} & \bf{0.9826} & \bf{47.974} & \bf{0.9934} \\
        2, 2 w/ integration  & \bf{33.426} & \bf{0.9086} & \bf{38.634} & \bf{0.9587} & 43.291      & \bf{0.9826} & 47.826      & 0.9933      \\
        \hline
    \end{tabular}
\end{table}

\begin{figure}[t!]
    \centering
        \raisebox{-0.25\height}{\subfigure[]{\includegraphics[width=22mm]{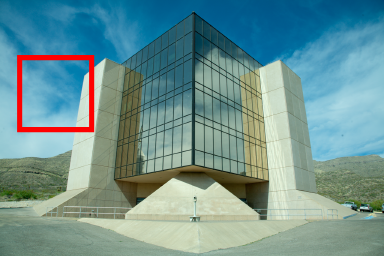}}}
        \begin{tabular}{cccc}
        \includegraphics[width=22mm]{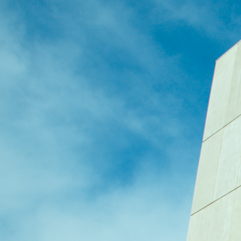} &
        \includegraphics[width=22mm]{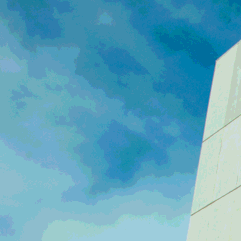} &
        \includegraphics[width=22mm]{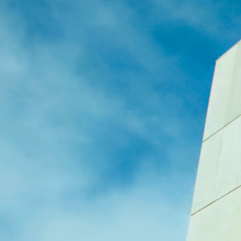} &
        \includegraphics[width=22mm]{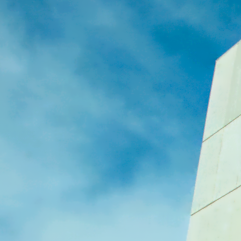} \\ 
        (b) & (c) & (d) & (e)\\ \addlinespace[2ex]
        \includegraphics[width=22mm]{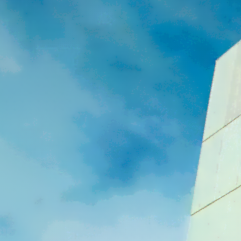} &
        \includegraphics[width=22mm]{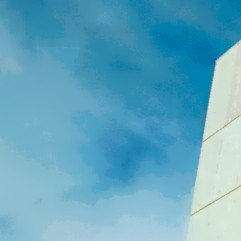} &
        \includegraphics[width=22mm]{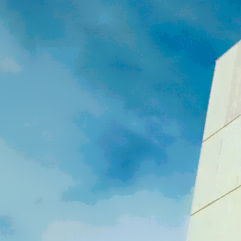} &
        \includegraphics[width=22mm]{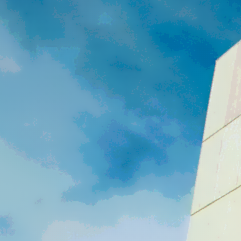} \\
        (f) & (g) & (h) & (i)\\
        \end{tabular}
    \caption{Comparison of BDE results with different number of integrated multi-scale features.
    (a) Original 8-bit image (b) Cropped 8-bit image (c) Input 3-bit image (d) 3-bit to 8-bit result of BitNet (e)$\sim$(i) Results from the single trained BitNet which additive connections are gradually removed from the smallest feature}
    \label{fig:multi3}
\end{figure}

Our novel multi-scale feature integration (MSFI) module in BitNet enlarges and integrates the previous features at the end of the upscaling. To evaluate how much MSFI improves the performance, we trained the network with and without MSFI and compared their performance. Figure \ref{fig:multi4} and Table \ref{table:integration} show the test results with the MIT-Abode FiveK dataset. Figure \ref{fig:multi4}(e) shows that BitNet with MSFI eliminates the false contours more effectively. In Table \ref{table:integration}, MSFI shows a noticeable performance improvement for 3 and 4-bit to 16-bit BDE. When bit-depth of the input is 5 or 6, the performance is slightly degraded, but the visual differences are minimal among the results. Additionally, we investigated the effect of MSFI by gradually discarding additive connections in MSFI from the smallest feature in our single trained BitNet and visualized each BDE result. As shown in Fig. \ref{fig:multi3}, more disconnections lead to worse false contour removal. Therefore, using enlarged features from small features is important for removing false contour artifacts.

\section{Conclusion}
In this paper, we use a learning-based approach for bit-depth expansion. We have carefully designed our BitNet with a novel multi-scale feature integration for effective false contour removal and investigated the impact of various combinations of the dilation rates. We have also found that, unlike traditional ways which separately handle each channel, it is better to treat all the color channels together for effective color restoration. Compared to other existing methods, our network estimates more accurate HBD images at a much faster speed. However, BitNet tends to depend on learned color patterns. This can be improved by combining a channel-wise version of BitNet or adopting random hue and contrast augmentations. In addition, the performance can be increased by using a well-designed loss function \cite{loss}.
\\
\\
\textbf{Acknowledgement.} This work was supported by MCST (Ministry of Culture, Sports\&Tourism)/KOCCA(KoreaCreativeContentAgency) (R2016030044 - Development of Centext-Based Sports Video Analysis, Summarization, and Retrieval Technologies). The authors would like to thank Jing Liu \cite{IPAD} for releasing source codes for various BDE methods.

\bibliographystyle{splncs04}
\bibliography{mybibliography}
\end{document}